\documentclass[showpacs,superscriptaddress,twocolumn,amsmath,aps]{revtex4}

\usepackage{graphicx}
\begin{document}
\draft

\title{Optimal control of molecular electronics by femtosecond laser
  pulses}

\author{GuangQi Li} 
 \affiliation{School of Engineering and Science, Jacobs University Bremen,
 Campus Ring 1, 28759 Bremen, Germany}
\author{Sven Welack} 
 \affiliation{Department of Chemistry, University of California, Irvine, California 92697-2025, USA}
\author{Michael  Schreiber}
\affiliation{Institut f\"ur Physik, Technische Universit\"at Chemnitz, 09107
  Chemnitz, Germany}    
\author{Ulrich Kleinekath\"ofer}
 \affiliation{School of Engineering and Science, Jacobs University Bremen,
 Campus Ring 1, 28759 Bremen, Germany}
 
\date{\today}

\begin{abstract}
  Combining the features of molecular wires and femtosecond laser pulses
  gives the unique opportunity to optically switch electron currents in
  molecular devices with very high speed. Based on a weak-coupling
  approximation between wire and leads a quantum master equation for the
  population dynamics and the electric current through the molecular wire
  has been developed which allows for arbitrary time-dependent laser fields
  interacting with the wire. This formalism is combined with the theory of
  optimal control. For a tight-binding approximation of the wire we show
  how to compute the laser pulses to switch the current through the wire on
  and off. With this approach the desired pattern of the current in time
  can be chosen in an almost arbitrary fashion.
\end{abstract}

\maketitle


The field of molecular electronics and especially of molecular wires has
attracted much interest recently, experimentally as well as theoretically
\cite{nitz03a,ghos04}.  In scattering theories one usually uses the
Landauer-B\"uttiker formula \cite{butt86} which was first developed for
steady-state situations and later extended to deal with time-dependent
fields \cite{meir92,wing93}, applying non-equilibrium Green's functions. In
a different class of approaches one treats the coupling of the wire to the
leads perturbatively \cite{kohl04a,li05,cui05,ovch05a,wela05a,harb06}.
Some of these treatments utilize knowledge from the field of dissipative
quantum dynamics and derive quantum master equations (QMEs) to describe the
current through a molecular wire.  But instead of coupling the system,
i.e.\ the wire, as usual to a bosonic heat bath it is now coupled to a
fermionic particle reservoir with which it can exchange particles. The
obtained QMEs easily allow us to apply external optical fields which
influence the dynamics in a direct manner by changing the wire part but
also in an indirect way by influencing the wire-lead coupling. For
monochromatic fields the Floquet theory has been employed
\cite{lehm02,tikh02b,lehm03a,kohl04a}. We use a technique in the field of
dissipative quantum dynamics in which a special parameterization of the
so-called spectral density of the reservoir leads to a set of coupled
equations for a primary and several auxiliary density matrices. This
formalism allows for an arbitrary time dependence of the laser pulses
\cite{meie99}. 

For cw-laser light it was shown \cite{lehm03a} that, within a simple
tight-binding model, the time-averaged current through a molecular wire
could be suppressed choosing the right amplitude of the field. This is
known as the phenomenon of coherent destruction of tunneling (CDT) and was
first discovered in two-level systems \cite{gros91}. Recently we
demonstrated \cite{klei06b,li07a} that one can also suppress the current
using short Gaussian laser pulses.  The maximum amplitude of these pulses
has to fulfill the same conditions as for a cw-laser pulse. Though the
current can be suppressed the shape of the current pattern in time cannot
be controlled easily.

Lately an optimal control theory (OCT) for time-independent targets has
been extended to control target states which are distributed in time
\cite{kais04,kais05} or which are time-dependent \cite{serb05}.  In this
Letter we apply the OCT to molecular wires with a small number of sites.
After selecting a time-dependent current target, one utilizes OCT to
optimize the time-dependent control field. Under the influence of this
field, indeed a current which is similar to the predefined current pattern
is obtained.  Although we concentrate on the topic of molecular wires here,
the techniques described below can, in principle, be applied to coupled
quantum dots as well.

The time-dependent Hamiltonian of the investigated molecular junction is
separated into the relevant system $H_S(t)$, mimicking the wire, and
reservoirs $H_R$ modeling the leads
\begin{equation} \label{equ:Ham_total}
H(t)=H_S(t)+H_F(t)+H_R+H_{SR}
\end{equation}
with wire-lead coupling $H_{SR}$.  
Denoting the creation (annihilation) operator at site $n$ by $c^\dagger_n$
($c_n$) the tight-binding description of the electrons in the molecular
wire reads
\begin{eqnarray} \label{equ:Ham_wire}
H_S(t)&=&\sum_n \varepsilon_n c_n^\dagger c_n - \Delta
\sum_{n} ( c_n^\dagger c_{n+1} +c_{n+1}^\dagger c_{n})~.
\end{eqnarray}
The first term describes the on-site energies and the second term the
nearest-neighbor hopping.  For simplicity electron spin and interaction
have been neglected since they will not qualitatively change the optimal
control scenario.  Denoting the dipole operator which will be detailed
below by $\mu{}$, the coupling between the wire and the laser field $E(t)$
reads
\begin{equation} \label{equ:Ham_laser}
H_F(t)= - \mu{}E(t)~.
\end{equation}

The environment of the wire consists of two electronic leads that are
modeled by two independent reservoirs of uncorrelated electrons in thermal
equilibrium. For each lead, the Hamiltonian $H_R$ is given by $
\label{equ:Ham_lead} H_R=\sum_q \omega_q b_q^\dagger b_q $ with
$b_q^\dagger$ creating and $b_q$ annihilating an electron in the
corresponding reservoir mode $\vert q \rangle$ with energy $\omega_q$.
We set $\hbar\equiv 1$. 
In further derivations we will only refer to the left lead but the
formalism has to be applied as well to the right lead coupled to the last
site $N$ of the wire.  The coupling of the left electronic lead with the
first site of the wire is given by
\begin{equation} \label{equ:Ham_coup}
H_{SR}= \sum_{x=1}^2 K_x  \Phi_x =  \sum_q (V_q c_1^\dagger b_q + V_q^* b_q^\dagger c_1)
\end{equation}
with $\Phi_1=\sum_q V_q b_q$, $\Phi_2=\sum_q V_q^* b_q^\dagger$,
$K_1=c_1^\dagger$, $K_2=c_1$, and a wire-lead coupling strength $V_q$ for
each reservoir mode.

Starting with the time-convolutionless approach, a time-local QME based on
a second-order perturbation theory in the molecule-lead coupling has been
developed for the reduced density matrix $\rho(t)$ of the molecule
\cite{wela05a,klei06b}
\begin{eqnarray}\label{equ:master2local}
\frac{\partial\rho(t)}{\partial t}= -i \mathcal L_S(t) \rho(t)
-i \mathcal L_F \rho(t)-D(t) \rho(t)~,
\end{eqnarray}
\begin{eqnarray}\label{equ:diss1}
  D(t) \rho(t)&=&\sum_{xx'} \left[ K_{x} \Lambda_{xx'}(t)\rho(t) -
    \Lambda_{xx'}(t)\rho(t)K_{x} \right. \nonumber \\
  && \left. - K_{x}\rho(t)\widehat \Lambda_{xx'}(t)+\rho(t)\widehat
    \Lambda_{xx'}(t)K_{x} \right]
\end{eqnarray}
with auxiliary operators for the wire-lead coupling
\begin{eqnarray}\label{equ:aux1local}
\Lambda_{xx'}^{}(t) &=& \int_{t_0}^t \mathrm dt' C_{xx'}(t-t')  
U_S(t,t')  K_{x'} \\ 
\widehat \Lambda_{xx'}(t)& =& \int_{t_0}^t \mathrm dt' C_{x'x}^*(t-t') 
 U_S(t,t')  K_{x'}~.
\end{eqnarray}
Here we employed the definitions $U_S(t,t')=T_+\exp\left\{-i \int_{t'}^t
  \mathrm d\tau \left( \mathcal L_S(\tau)+\mathcal L_F(\tau) \right)\right\}$,
$\mathcal L_S(\tau) =[H_S(\tau),\bullet]$, $\mathcal L_F(\tau) 
=[H_F(\tau),\bullet]$ with the time-ordering operator $T_+$ and the reservoir
correlation functions $C_{x x'}(t)$.  For later convenience we also
introduce $\mathcal L_\mu{} =[\mu{} ,\bullet]$.
Using the electron number operator $N_l=\sum_q c_q^{\dagger} c_q$ of the
left lead with the summation performed over the reservoir degrees of
freedom yields
\begin{eqnarray}\label{equ:current}
I_l(t)=e\frac{\mathrm d}{\mathrm dt} \mathrm{tr} \, \lbrace N_l \rho(t)
\rbrace = \mathrm{tr} \left\{ \mathcal I_l(t) \rho(t) \right\}  
\end{eqnarray}
with the current operator
\begin{eqnarray}\label{equ:currentoperator}
  \mathcal I_l(t)\rho(t) &=& e \left[ c_{1}^\dagger
    \Lambda_{12}(t)\rho(t) - c_{1}^\dagger \rho(t)
    \widehat\Lambda_{12}(t)  \right. \nonumber \\ 
  &&\left .+ c_{1}\rho(t) \widehat\Lambda_{21}(t)
    -c_{1}\Lambda_{21}(t)\rho(t)  \right]
\end{eqnarray}
and the elementary charge $e$. This equation describes the current
$I_l(t)$ from the left lead into the molecule. A similar expression holds
for $I_r(t)$ from the right lead into the molecule.  As in a steady-state
situation, the net transient current can be defined as
$I(t)=(I_l(t)-I_r(t))/2$.  In addition, we define the net current
operator $\mathcal I(t)=(\mathcal I_l(t) -\mathcal I_r(t) )/2$.

Aim of this study is to determine laser pulses that result in a predefined
effect on the current through the molecular wire. This is achieved by
extremizing a control functional. Using techniques developed previously for
time-dependent targets \cite{kais06,koch04} we define as part of the
functional the difference between a preselected current pattern $P(t)$ and
the current obtained from the QME
\begin{equation}\label{equ:j0a}
  J_0(E)=\int \mathrm dt \,  \left\{ P(t)  -  \mathrm{tr} \left[ \mathcal I(t) \rho(t) \right] \right\} ^2 .
\end{equation}
To this we add a second part which shall ensure convergence
\cite{pala03,koch04,xu04}
\begin{equation}\label{equ:jtfa}
  J(E)=J_0(E)+\frac{\lambda}{2}\int\limits_{t_0}^{t_f} \mathrm d t \, \frac{\left(E(t)-\tilde{E}(t)\right)^2}{s(t)}
\end{equation}
with $\tilde{E}(t)$ being the laser field of the previous iteration step.
The penalty parameter $\lambda$ is a Lagrange multiplier.
Introducing the time-dependent function $s(t)$ we avoid a sudden switch-on
and switch-off behavior of the control field at the beginning and the end
of the propagation time.  In the present study either a Gaussian or a
squared sine function has been employed. The numerical results are rather
independent of the actual choice.

In order to calculate the extremum of the functional (\ref{equ:jtfa}),
the  functional derivative has to vanish \cite{kais06}:
\begin{equation}\label{equ:deri}
  \frac{\delta J(E)}{\delta E(t)}=0
  =\frac{\delta J_0(E)}{\delta E(t)}+\frac{\lambda}{2}\frac{\delta}{\delta
    E(t)}\int \mathrm d t
 \frac{(E(t)-\tilde{E}(t))^2}{s(t)}.
\end{equation}
Using the time-evolution operator $U(t,t',E)= T_+ \exp \left\lbrace
  -i\int_{t'}^{t} d \tau  \left[\mathcal L_S(\tau) + \mathcal L_F(\tau)- i
    D(\tau) \right] \right\rbrace$ together with $\rho(t)=
U(t,t_0,E)\rho(t_0)$, one gets
\begin{eqnarray}\label{equ:field}
E(t)&=&\tilde{E}(t)-\frac{s(t)}{\lambda} \mathrm{tr}_S \left\lbrace \chi(t) \left [ i \mathcal L_\mu{}\rho(t) - 
\frac{\delta D(t)}{\delta E(t)} \rho(t)\right ] \right\rbrace  \nonumber \\
&&-\frac{s(t)}{\lambda} \mathrm{tr} \left \lbrace  \frac{\delta \mathcal
  O(t)}{\delta E(t)} \rho(t) \right\rbrace 
\end{eqnarray}
with the operators
\begin{eqnarray}\label{equ:zmda}
  \chi(t)&=&\int_t^{t_f} \mathrm d\tau \, \mathcal O(\tau) U(\tau,t,E) , \\ 
  \mathcal O(\tau)&=&2\left\lbrace  \mathrm{tr} \left[\mathcal I(\tau) \rho(\tau) \right] - P(\tau) \right\rbrace  \mathcal I(\tau)~.
\end{eqnarray}
The functional derivatives $\delta D/\delta E$ and $\delta \mathcal
O/\delta E$ are straightforward to evaluate but lengthy and will be given
elsewhere. For the  operator $\chi(t)$ one can derive a QME
\begin{eqnarray}\label{equ:chi}
  \frac{\partial\chi(t)}{\partial t}=-i \mathcal L_S(t) \chi(t)-i\mathcal L_F(t) \chi(t)+\overline D(t)\chi(t) -\mathcal O(t)
\end{eqnarray}
where the inhomogeneous term $\mathcal O(t)$ reflects the time-distributed
target state. The wire-lead coupling operator $\overline D(t)$ in this QME is
given by
\begin{eqnarray}\label{equ:diss2}
  \overline D(t)\chi(t)&=&\sum_{xx'}\left[
 \chi(t) K_{x} \Lambda_{xx'}(t) - K_{x}\chi(t)\Lambda_{xx'}(t)
 \right. \nonumber \\ 
  && \left.     - \widehat \Lambda_{xx'}(t)\chi(t)K_{x}
 + \widehat \Lambda_{xx'}(t)K_{x}\chi(t)
      \right]~.
\end{eqnarray}
While Eq.~(\ref{equ:master2local}) is propagated forward in time,
Eq.~(\ref{equ:chi}) is propagated backward in time. These two equations can
now be solved in an iterative fashion.  One starts with an equilibrated
state on the wire $\rho(0)$ at time $t_0=0$ which is unchanged during the
iterative process and chooses an initial guess for the field $E(t)$.  The
density matrix $\rho(t)$ is propagated forward from initial time $t=t_0$ to
final time $t=t_f$ using $E(t)$.  In a second step, one obtains a new field
$E(t)$ from Eq.~(\ref{equ:field}), in order to propagate the target
operator $\chi(t)$ backward from final time $t=t_f$ to initial time
$t=t_0$.  One stores the field $E(t)$ as $\tilde{E}(t)$ for the next step,
in which we again calculate $E(t)$ and propagate $\rho(t)$ from $t_0$ to
$t_f$ with the initial condition $\rho(t_0)$. This scheme is repeated
iteratively until convergence. As discussed in Ref.~\onlinecite{serb05}
some of the density matrices and fields can be stored during the iterative
process to save computational effort.

For the calculations shown below we consider a wire consisting of two sites
without spin, equal site energies $\varepsilon_1=\varepsilon_2$, intersite
coupling $\Delta=0.1$ eV, and a dipole operator 
$\mu{}= (c_1^\dagger c_1 -c_2^\dagger c_2)/2$.
A bias $V_b=0.4$ eV symmetric with respect to the site energies
$\varepsilon_n$ determines the Fermi energies
$\varepsilon_{F,l}=\varepsilon_1+V_b/2$ and
$\varepsilon_{F,r}=\varepsilon_2-V_b/2$ which describe the occupation of
the reservoir modes. The coupling between the wire and the leads is almost
in the wide-band limit and the maximum coupling strength is 0.1 $\Delta$.
For the first and third example below $\lambda$ was set to 0.003 and for
the second calculation to 0.0006.

\begin{figure}
\centerline{
\includegraphics[width=7.cm,clip]{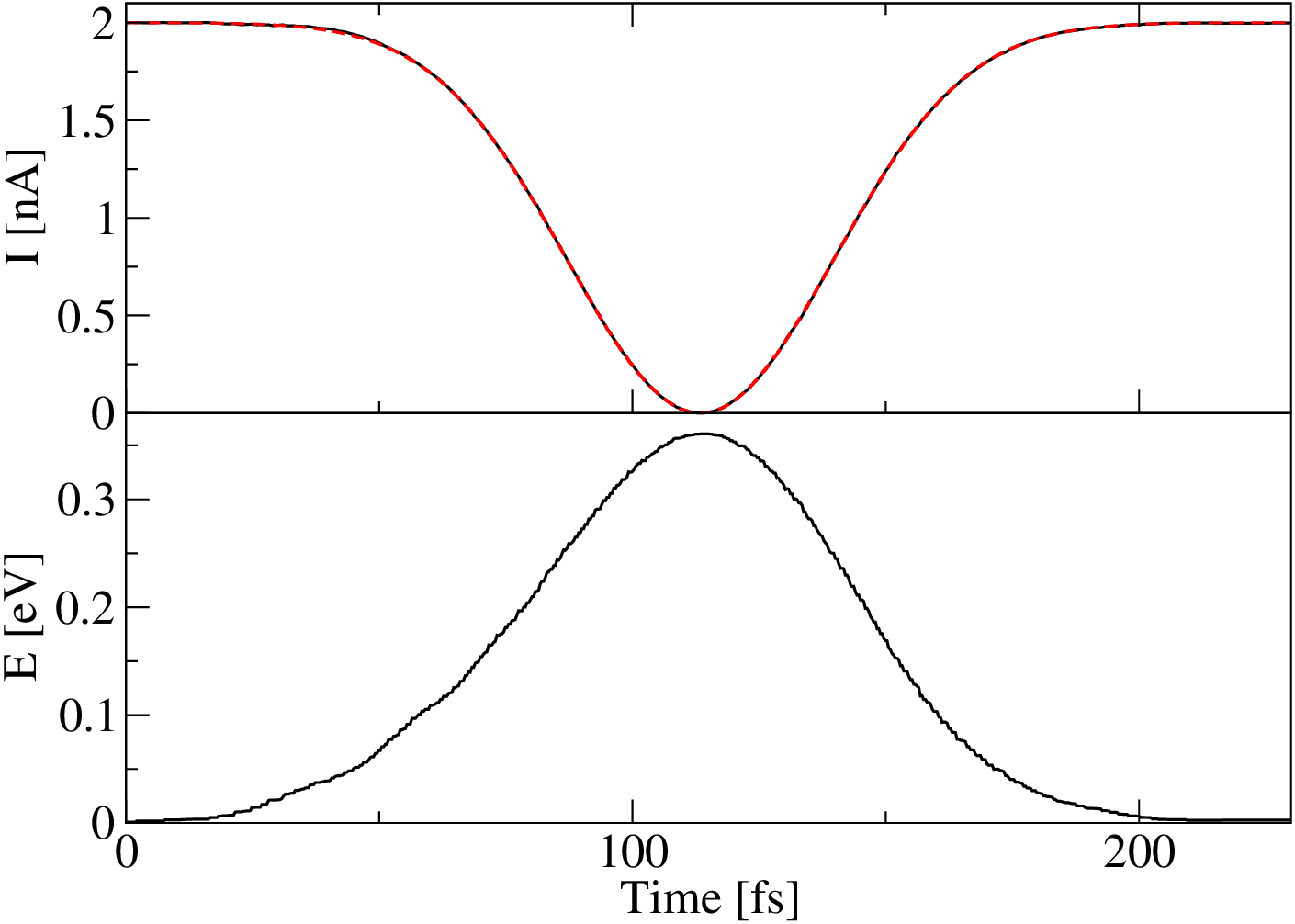}}
\caption{Target current $P(t)$ (dashed) and calculated current $I(t)$
  (solid) in the upper panel. The control field $E(t)$ is shown in the
  lower panel.}
\label{f.0}
\end{figure}

As a first simple example of current control we consider a case in which
the current is initially constant and the goal is to suppress the current
following a Gaussian shape in time.  In Fig.~\ref{f.0} the current target
$P(t)$ is shown together with the obtained current and the corresponding
laser field $E(t)$. Until the initial time $t_0$ = 0  the laser field
is turned off and the system is equilibrated, i.e.\ in a steady state
leading to a time-independent current.  As can be seen the optimal control
current matches the target very well.  The optimal laser field is slightly
asymmetric and its maximum occurs a few femtoseconds before the minimum of
the current.  This is in agreement with the earlier observation
\cite{klei06b,li07a} that a Gaussian shaped envelope of the laser field
leads to a slightly asymmetric current pattern with a minimum delayed by a
few femtoseconds.

\begin{figure}
\centerline{
\includegraphics[width=7.cm,clip]{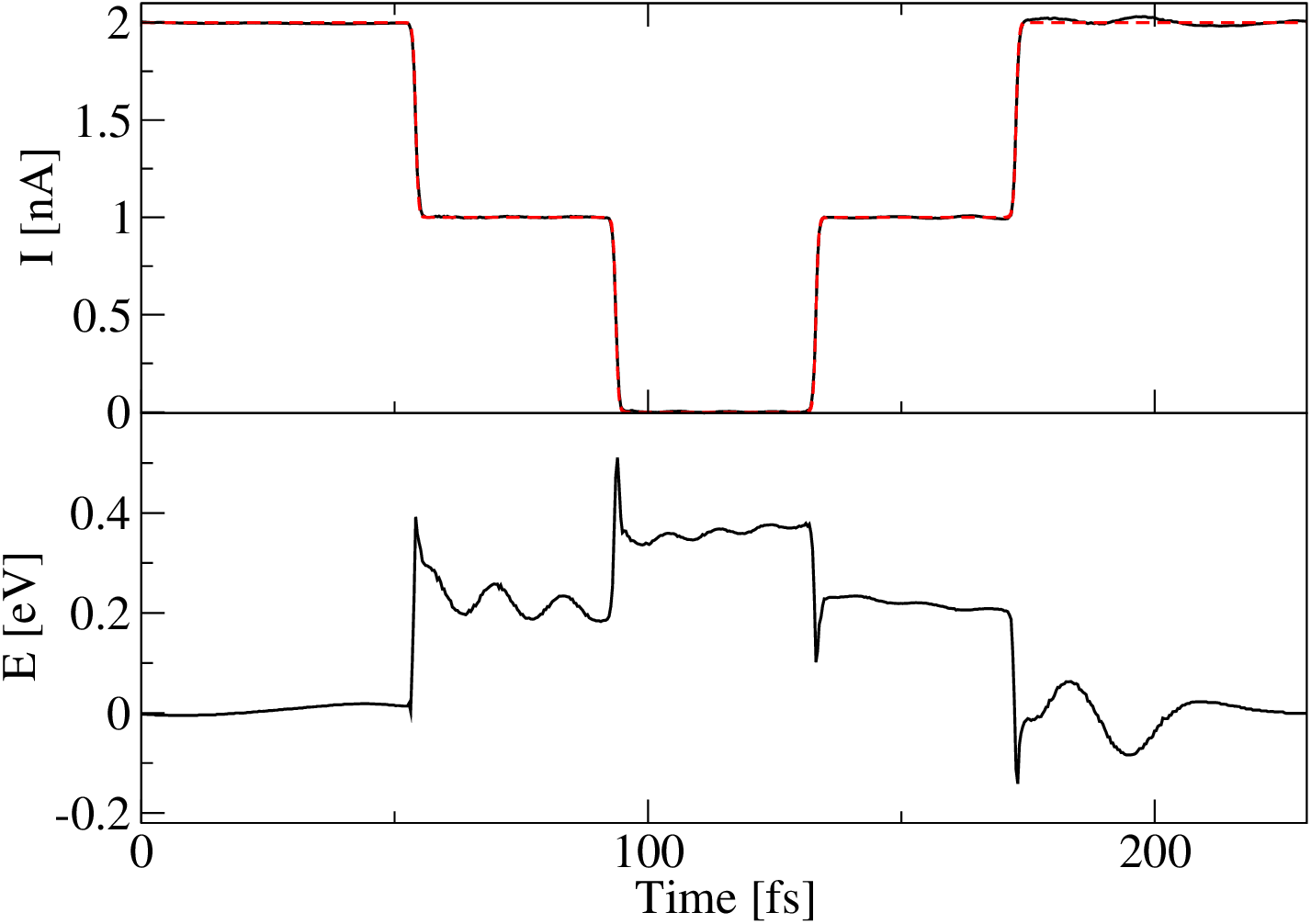}}
\caption{Same as in Fig.~1 for a different control target.}
\label{f.1}
\end{figure}

In the second example the complexity of the control task is increased.  As
shown in Fig. \ref{f.1} the current target $P(t)$ is a symmetric double
step function and the goal is achieved by the optimal control algorithm
rather accurately. Only at the last step there is some visible deviation
between target and achieved current. The optimized laser field in the lower
panel of Fig.~\ref{f.1} also shows step-like features but additionally
oscillations and peaks. In contrast to the previous control target, the
rapidly changing target current pattern requires larger changes in the
electric field. To compensate the effect of these large peaks on the
current at later times, additional oscillations of the field are needed to
achieve constant values of the current at the plateaus.

\begin{figure}
\centerline{
\includegraphics[width=7.cm,clip]{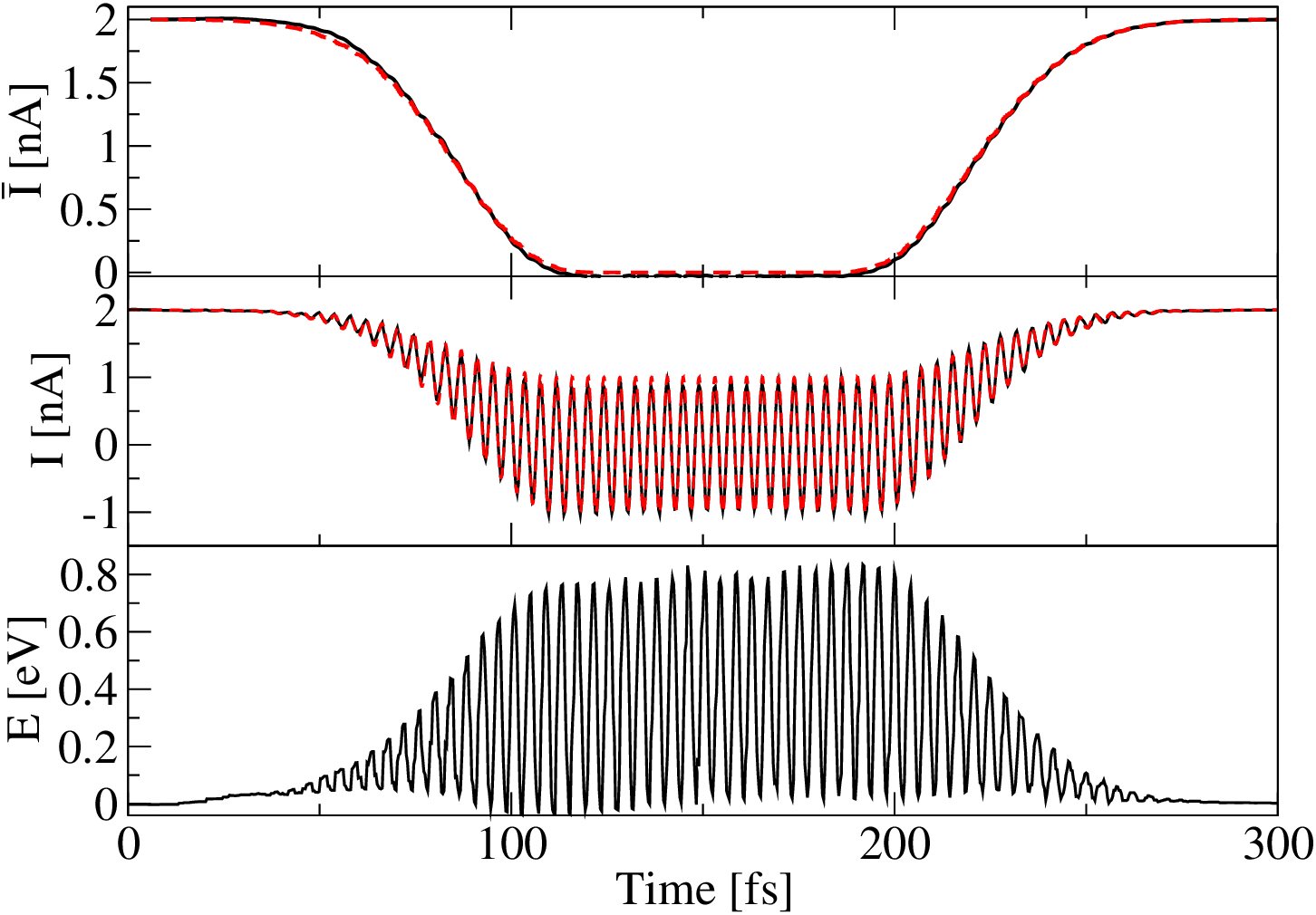}}
\caption{A control target including a highly oscillating pattern. The
  control target (dashed) and the achieved current (solid) are displayed in
  the middle panel while in the top panel an average over five oscillatory
  periods is shown.  The control field is given in the lowest part of the
  figure.}
\label{f.2}
\end{figure}

In many experimental setups the laser field has a high carrier frequency
and a more slowly varying envelope function. To mimic such a scenario we
created a target for the current which consists of a fast oscillating
pattern with $\omega=1$ eV modulated by a slowly varying envelope function,
namely a current depletion of half-Gaussian form, a constant part, and
again a half-Gaussian shape for the increasing current. This pattern
together with the current achieved by the optimal control algorithm is
shown in the middle panel of Fig.~\ref{f.2}. Despite the rapidly varying
target function the goal is accurately achieved. Similar to previous
studies of CDT we also calculated a mean current which is determined by
averaging the current over a few cycles of the carrier frequency, here 5
periods. This procedure can of course be performed for the target as well
as for the shaped current and the results are shown in the top panel of
Fig.~\ref{f.2}. As can be seen the average current is suppressed nicely in
such a scenario.

In previous studies \cite{lehm03a,klei06b,li07a} CDT was utilized to
suppress the current which is possible for certain values of the amplitude
$A$ of the laser field: The fraction $A/\omega$ has to be equal to a zero of
the zeroth-order Bessel function. In the present case the maximal value of
the field is about $A/\omega=0.8$ which is certainly far off a zero of the
zeroth-order Bessel function. Therefore the present current depletion
cannot be explained by CDT.

\begin{figure}
\centerline{
\includegraphics[width=6.5cm,clip]{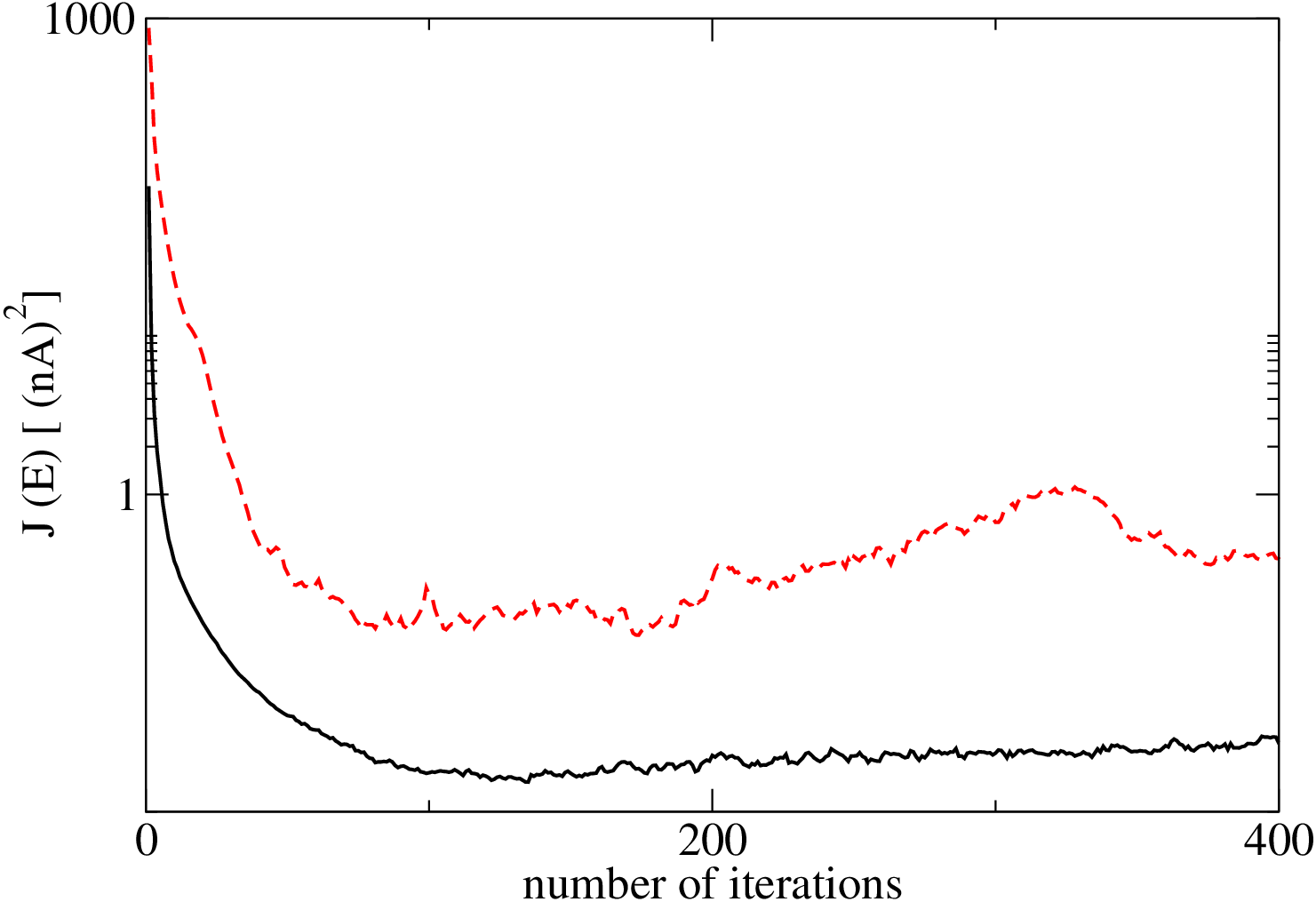}}
\caption{ The value of the control function $J(E)$. 
  Solid for the step function as shown in Fig.~\ref{f.1} and dashed
  for the symmetric Gaussian shape in Fig.~\ref{f.2}. }
\label{f.3}
\end{figure}

The convergence of the iterative control algorithm for the second and third
example is displayed in Fig.~\ref{f.3}. It shows how the value of $J$
decreases with increasing number of iterations. For a simple target like a
step function, $J$ decreases very fast and then gets more or less constant.
In contrast for a complex target with a fast oscillating function, it
decreases not so fast in the beginning and later even increases slightly.
To further decrease the control function one could increase the Langrange
parameter $\lambda$ for large iteration numbers.


In conclusion, we have demonstrated that it is theoretically possible to
achieve a time-dependent pattern of the current through a molecular
junction. The optimized current can be obtained using an optimal control
field. Although we used a simplified model for a molecular wire the present
investigation shows that it is worthwhile to study the combination of
molecular wires and optimal control theory on the femtosecond time scale,
theoretically as well as experimentally.  In experiment feedback coherent
control for complex systems has been successful \cite{brix03} so that an
experimental implementation of the ideas described above might be feasible
in the future.

The authors would like to thank the Deutsche Forschungsgemeinschaft for
financial support within SPP 1243.

\end{document}